\documentclass[english,12pt]{article}
\usepackage{array}
\usepackage{graphicx}
\usepackage{amssymb}
\usepackage{amsmath}
\usepackage{multirow}
\usepackage{prettyref}
\usepackage{babel}
\usepackage{units}
\usepackage[latin1]{inputenc}
\usepackage{amsfonts}
\usepackage{amssymb}
\usepackage{babel}
\usepackage{cite}
\def\@fmsl@sh#1#2#3{\m@th\ooalign{$\hfil#1\mkern#2/\hfil$\crcr$#1#3$}}
 \def\eq#1\en{\begin{equation}#1\end{equation}}
\def\s[#1,#2]{[#1\stackrel{\star}{,}#2]}
\def\sx[#1,#2]{[#1\stackrel{\star_{x}}{,}#2]}

\newcommand{\nc}{\newcommand}
\nc{\beq}{\begin{equation}}
\nc{\eeq}{\end{equation}}
\nc{\beqa}{\begin{eqnarray}}
\nc{\eeqa}{\end{eqnarray}}

\def\bc{\begin{center}}
\def\ec{\end{center}}

\def\to{\rightarrow}

\def\gsim{\mathrel{\mathpalette\atversim>}}

\def\bc{\begin{center}}
\def\ec{\end{center}}

\def\gsim{\mathrel{\rlap{\lower4pt\hbox{\hskip1pt$\sim$}}

    \raise1pt\hbox{$>$}}}       

\def\gsim{\mathrel{\rlap{\lower4pt\hbox{\hskip1pt$\sim$}}
    \raise1pt\hbox{$>$}}}       

\begin{document}
\makeatletter
\def\fmslash{\@ifnextchar[{\fmsl@sh}{\fmsl@sh[0mu]}}
\def\fmsl@sh[#1]#2{%
  \mathchoice
    {\@fmsl@sh\displaystyle{#1}{#2}}%
    {\@fmsl@sh\textstyle{#1}{#2}}%
    {\@fmsl@sh\scriptstyle{#1}{#2}}%
    {\@fmsl@sh\scriptscriptstyle{#1}{#2}}}
\def\@fmsl@sh#1#2#3{\m@th\ooalign{$\hfil#1\mkern#2/\hfil$\crcr$#1#3$}}
\makeatother

\thispagestyle{empty}
\begin{titlepage}
\boldmath
\begin{center}
  \Large {\bf Bounds on the non-minimal coupling of the Higgs boson to gravity}
    \end{center}
\unboldmath
\vspace{0.2cm}
\begin{center}
{  {\large Michael Atkins}\footnote{m.atkins@sussex.ac.uk} and {\large Xavier Calmet}\footnote{x.calmet@sussex.ac.uk} }
 \end{center}
\begin{center}
{\sl Physics $\&$ Astronomy, 
University of Sussex,   Falmer, Brighton, BN1 9QH, UK 
}
\end{center}
\vspace{5cm}
\begin{abstract}
\noindent
We derive the first bound on the value of the Higgs boson non-minimal coupling to the Ricci scalar. We show that the recent discovery of the Higgs boson at the Large Hadron Collider at CERN implies that the non-minimal coupling is smaller than $2.6\times 10^{15}$. 
\end{abstract}  
\end{titlepage}



\newpage
 Einstein's theory of general relativity \cite{Einstein:1916vd} provides a complete description of gravity on length scales that range from fractions of a millimeter to the cosmological horizon. It has been verified in numerous experiments and no deviations from the theory have been found. Despite this extraordinary success on macroscopic scales, general relativity cannot be the correct theory at the quantum level as it is not renormalizable. In other words, Einstein's gravity will break down at energies approaching the Planck scale, where a complete theory of quantum gravity is expected to replace it. General relativity should thus be regarded as simply the leading order term in an effective field theory description of a more fundamental high energy theory. Any quantum theory of gravity which is diffeomorphism invariant can be described at  energies below the reduced Planck  scale $M_P = 2.435 \times 10^{18}$ GeV  by an effective theory, see e.g. \cite{Donoghue:1994dn,Donoghue:2012zc}, whose leading order terms are given by
\begin{eqnarray}\label{action1}
S = \int d^4x \, \sqrt{-g} \left[ \frac{M_P^2}{2} (\mathcal{R}-2 \Lambda) + c_1 \mathcal{R}^2 + c_2 \mathcal{R}_{\mu\nu}\mathcal{R}^{\mu\nu}  + \mathcal{O}(M_P^{-2}) \right] 
\end{eqnarray}
where $\Lambda$ is the cosmological constant, $\mathcal{R}$ and $\mathcal{R}_{\mu\nu}$ are the Ricci scalar and tensor respectively. The Wilson coefficients $c_1$ and $c_2$ are dimensionless parameters that encode the strength of the contribution of the related operators. This effective theory is an expansion in the curvature, and as such, the next to leading order terms are the dimension four curvature squared operators. Higher dimensional operators are suppressed by increasing powers of the Planck scale $M_P$ and will thus be even less significant at low energies. In principle the Wilson coefficients could be calculated within a specific theory of quantum gravity by matching the effective theory to the full underlying theory. Measuring these coefficients could thus allow to differentiate between different theories of quantum gravity.
 It has been shown by Stelle in 1977 \cite{Stelle:1977ry} that the terms  $c_1 R^2$ and $c_2 R^{\mu\nu}R_{\mu\nu}$ lead to Yukawa-like corrections to the Newtonian potential of a point mass $m$: 
\begin{eqnarray}
\Phi(r) = -\frac{Gm}{r} \left( 1+\frac{1}{3}e^{-m_0 r}-\frac{4}{3}e^{-m_2 r} \right) 
\end{eqnarray}
where $G$ is Newton constant,
\begin{eqnarray}
m_0^{\,-1}=\sqrt{32\pi G \left(3c_1-c_2 \right)}
\end{eqnarray}
and
\begin{eqnarray}
m_2^{\,-1}=\sqrt{16\pi G c_2}.
\end{eqnarray}
 Using recent experimental advances \cite{Hoyle:2004cw}, one finds that the coefficients $c_1$ and $c_2$ are  constrained to be less than $10^{61}$ \cite{Calmet:2008tn} in the absence of accidental fine cancellations between both Yukawa terms. Attempts to bound these terms using astrophysical measurements have been reviewed in \cite{Psaltis:2008bb}.

While the true nature of quantum gravity is still unclear, the standard model of particle physics represents a fully consistent quantum field theory description of the observed particles of nature that has met all experimental tests to date. Indeed, recent results from the CERN Large Hadron Collider (LHC) reveal the existence of a new particle consistent with the standard model Higgs boson at a mass of approximately 125 GeV  \cite{:2012gk,:2012gu}. Assuming that the particle is confirmed to be the Higgs boson, the discovery completes our observation of all the ingredients of the standard model.

With the discovery of the Higgs boson, there is one additional dimension four operator that should be present in the effective theory: the non-minimal coupling between the Higgs field and gravity, 
\begin{equation}\label{nonMin}
S \supset \int d^4x \,\sqrt{-g} \, \xi H^\dagger H \mathcal{R},
\end{equation}
where $H$ represents the Higgs doublet. The dimensionless non-minimal coupling $\xi$ is so far an unconstrained parameter of nature that remains to be measured. Note that this new operator is leading order in the expansion in the curvature. This non-minimal coupling could play an important role in cosmological models \cite{Frommert:1996gu} including inflationary scenarios, see e.g. \cite{CervantesCota:1995tz,Bezrukov:2007ep}. The possibility of a non-minimal coupling between a scalar field and the Ricci curvature was first noted in reference \cite{Chernikov:1968zm}\footnote{We thank an anonymous referee for bringing this paper to our attention.} and it was later pointed out that this coupling will be generated when one quantizes a scalar field on a curved spacetime \cite{Callan:1970ze}.

The aim of this letter is to derive the first known bounds on the size of the coupling $\xi$. The main approach is to make use of a decoupling between the physical Higgs boson and the rest of the standard model particles that accompanies a large non-minimal coupling and study the effect this would have on the production and decay of the Higgs boson at the LHC. We also estimate the expected reach of future high energy, high luminosity runs at the LHC and proposed International Linear Collider (ILC) to improve the bounds on $\xi$. Finally we add some comments on Higgs boson decays to gravitons, the effect on the Higgs boson's mass of a large non-minimal coupling and the consequences of our results for various models found in the literature.

 Including the non-minimal coupling and the standard model Lagrangian $\mathcal{L}_{SM}$, the action (\ref{action1}) becomes
\begin{eqnarray}
\label{action2}
S = \int d^4x \, \sqrt{-g} \left[ \left( \frac12  M^2 + \xi H^\dagger H \right) \mathcal{R} - (D^\mu H)^\dagger (D_\mu H) + \mathcal{L}_{SM}  \right] 
\end{eqnarray}
where we have suppressed the cosmological constant term, the curvature squared terms and higher dimensional operators, and  we have replaced the Planck scale with a generic mass scale to be fixed below. We have also explicitly written the kinetic term for the Higgs field, which is normally contained in $\mathcal{L}_{SM}$. After electroweak symmetry breaking, the Higgs boson gains a non-zero vacuum expectation value, $v=246$ GeV, $M$ and $\xi$ are then fixed by the relation
\begin{eqnarray}
\label{effPlanck}(M^2+\xi v^2)=M_P^2 \, .
\end{eqnarray}
From this it is clear that $\xi \le M_P^2/v^2 \simeq 10^{32}$. Note that $\xi$ can be of arbitrary size if negative. One might naively expect that if $|\xi|$ is much below $10^{32}$ then its effects would not be observable in low energy experiments. This however turns out to be false as we will now show.

The easiest way to see the decoupling effect of the Higgs boson is to make a transformation to the Einstein frame\footnote{This effect was first realized for the Higgs boson in a paper by Van der Bij \cite{vanderBij:1993hx} where it was assumed that $M=0$ and the Planck scale is generated entirely by the Higgs boson's vacuum expectation value with $\xi \simeq 10^{32}$. An earlier reference to the same effect in grand unified theories was made by Zee \cite{Zee:1978wi} where he assumed the Higgs boson's vacuum expectation value that breaks the Grand Unified Theory gauge symmetry could dynamically generate the Planck scale. See also \cite{Minkowski:1977aj} and references in \cite{Adler:1982ri}, where the Planck scale is generated via a symmetry breaking mechanism.}, $\tilde g_{\mu\nu}=\Omega^2 g_{\mu\nu}$, where $\Omega^2 =(M^2+2\xi H^\dagger H)/M_P^2$. The action in the Einstein frame then reads
\begin{eqnarray}
\label{actionEins}
S=\int d^4x \, \sqrt{- \tilde g} \left[\frac{1}{2}M_P^2 \mathcal{\tilde R} -\frac{3 \xi^2}{M_P^2 \Omega^2}\partial^\mu (H^\dagger H ) \partial_\mu (H^\dagger H ) -\frac{1}{\Omega^2} (D^\mu H)^\dagger (D_\mu H) + \frac{\mathcal{L}_{SM}}{\Omega^4}  \right] \, .\end{eqnarray}
Expanding around the Higgs boson's vacuum expectation value and specializing to unitary gauge, $H=\frac{1}{\sqrt{2}}(0,\phi + v)^\top$, we see that in order to have a canonically normalized kinetic term for the physical Higgs boson we need to transform to a new field $\chi$ where
\begin{equation}
\label{chiphi}
\frac{d\chi}{d\phi}=\sqrt{\frac{1}{\Omega^2}+\frac{6\xi^2 v^2}{M_P^2 \Omega^4}}\, .
\end{equation}
Expanding $1/\Omega$, we see at leading order the field redefinition simply has the effect of a wave function renormalization of $\chi = \phi/\sqrt{1+\beta}$ where $\beta =6 \xi^2 v^2 / M_P^2$. As a result, the Higgs boson's couplings to all the standard model particles get suppressed. For $\xi^2 \gg M_P^2 / v^2 \simeq 10^{32}$ the Higgs boson effectively decouples from the rest of the standard model.

This effect can also be understood in the original Jordan frame action (\ref{action2}) as arising from a mixing between the kinetic terms of the Higgs and gravity sectors. After fully expanding the Higgs boson around its vacuum expectation value and also the metric around a fixed background, $g_{\mu\nu}=\gamma_{\mu\nu}+h_{\mu\nu}/M_P$, we find a term proportional to $\frac{\xi v}{M_P} \phi \square h^\mu_\mu$. After correctly diagonalizing the kinetic terms and canonically normalizing the Higgs field we again find the physical Higgs boson gets renormalized by a factor $1/\sqrt{1+\beta}$.

At the LHC, the Higgs boson production and decay will be effected by the above suppression. At each vertex involving the Higgs boson coupled to standard model particles, a factor of $1/\sqrt{1+\beta}$ will be introduced. Clearly if $\beta \gg 1$ the Higgs boson would simply not be produced in a large enough abundance to be observed. 

In the following we will make the assumption that the Higgs boson like particle recently observed at the LHC is the standard model Higgs boson and that there are no other degrees of freedom beyond those present in the standard model and Einstein gravity. We will refer to the usual standard model total cross section for Higgs boson production and decay with $\beta=0$ as $\sigma_{\rm{SM}}$. If the cross section including a non-zero $\beta$ is given by $\sigma$, we are interested in the ratio $\sigma / \sigma_{\rm{SM}}$. The LHC experiments produce fits to the data assuming that all Higgs boson couplings are modified by a single parameter $\kappa$ \cite{LHCHiggsCrossSectionWorkingGroup:2012nn} which in our model corresponds to $\kappa=1/\sqrt{1+\beta}$. Using the narrow width approximation, the cross section for Higgs production and decay from any initial $i$ to final state $f$ is given by
\begin{equation}\label{crosssect}
\sigma(ii \to H) \cdot  {\rm BR}(H \to ff)= \sigma_{\rm SM}(ii \to H) \cdot  {\rm BR}_{\rm SM}(H \to ff) \cdot \kappa^2
\, .
\end{equation}
One might naively expect the cross section to be proportional to $\kappa^4$, but in the narrow width approximation this is not the case. The presence of the branching fraction, which is independent of a universal suppression of the couplings, leads to the cross section being proportional to $\kappa^2$. For a $125$ GeV Higgs the narrow width limit is an excellent approximation and is used in the determination of the signal strength at the LHC.

The ATLAS detector has currently measured the global signal strength $\mu=\sigma/\sigma_{\rm{SM}}=1.4 \pm 0.3$ \cite{:2012gk} and CMS has measured this as $\mu= 0.87\pm0.23$ \cite{:2012gu}. Combining these results gives $\mu=1.07\pm0.18$. This excludes $| \xi | >2.6 \times 10^{15}$ at the $95 \%$ C.L.

Reference \cite{Peskin:2012we} estimates the expected reach in the accuracy of the measurement of the Higgs boson couplings in a large number of processes in future runs at the LHC and the proposed ILC. We combine these results to give an estimated uncertainty in the global signal strength $\mu$. We assume a central value of $\mu = 1$. At a 14 TeV LHC with an integrated luminosity of 300 fb$^{-1}$, the uncertainty in the measurement of $\mu$ is expected to be 0.07 which would lead to a bound on $|\xi|<1.6 \times 10^{15}$. At the ILC with a center of mass of 500 GeV and an integrated luminosity of 500 fb$^{-1}$, the expected uncertainty on $\mu$ is 0.005, which gives a bound of $|\xi| < 4 \times 10^{14}$. Despite expected measurements of the total cross section to an accuracy better than $1\%$ at future high energy runs at the ILC, we cannot expect to push the constraints on $|\xi|$ below about $10^{14}$.

Given a large non-minimal coupling to gravity, one might also expect to have decreased observable rates for Higgs decays at the LHC arising from unobserved decays to gravitons. The effect is in fact very small as we will now discuss. The lowest order vertex in $\xi$ is a three point vertex connecting a single graviton line to two Higgs boson lines. This could introduce the possibility of a Higgs boson radiating off a single graviton before decaying to standard model particles. While this process is kinematically allowed for an off shell Higgs boson, it turns out that due to the nature of the derivative coupling, the amplitude for this process is always proportional the four-momentum squared of the emitted graviton and is therefore zero. There is no vertex allowing for a Higgs boson decaying to two gravitons after the kinetic terms have been properly normalized.  All other higher order processes will involve multiple Higgs bosons and as such will be extremely rare at any future collider. We conclude that the decoupling effect is the primary method available at colliders to put constraints on $\xi$. It would be of great interest if any cosmological or astrophysical effects were found that could compete with the constraints we have derived here.

We would like to make a short comment on the effect of the wave function renormalization on the Higgs boson self coupling. Clearly the wave function renormalization will also act to reduce the mass of the Higgs boson. This effect would have to be compensated by an increase in the Higgs boson self coupling. The increased self coupling would unfortunately not show up in direct searches attempting to measure the four point Higgs boson vertex since this will be further suppressed by a factor of $1/(1+\beta)$ coming from the additional two Higgs boson lines.

Over the years there has been considerable interest in the Higgs boson non-minimal coupling to gravity in the literature. This coupling is particularly important in models of ``induced gravity'' where the scale scale is generated spontaneously by setting $M=0$ and requiring that $\xi \simeq 10^{32}$ \cite{Frommert:1996gu,vanderBij:1993hx,Dehnen:1989vg}. Such a setup was also shown to be able to produce good inflation with the standard model Higgs boson acting as the inflaton \cite{CervantesCota:1995tz}. Clearly the discovery of the Higgs boson rules out such models on the grounds that with such a large $\xi$ the Higgs boson  would be almost completely decoupled from the rest of the standard model and would never be produced at a collider. In fact the decoupling effect for the Higgs boson we have used here was first observed in reference \cite{vanderBij:1993hx}. Later models of Higgs inflation used a much smaller value of the non-minimal coupling of the order of $10^4$ \cite{Bezrukov:2007ep}. Our results imply that colliders will not be able to probe the size of the non-minimal coupling down to these scales in the foreseeable future.

In conclusion, in this letter, we have set the first ever bound on the value of the Higgs boson's non-minimal coupling to the Ricci scalar which was the only remaining unconstrained dimension four operator in the effective theory obtained by coupling the standard model to general relativity. We have shown that the recent discovery of the Higgs boson at the Large Hadron Collider at CERN implies that the absolute value of the non-minimal coupling is smaller than $2.6\times 10^{15}$.  We consider that this bound is a step  on our path to a better understanding of quantum gravity and probing quantum gravity experimentally.

{\it Acknowledgments:} We would like to thank J. No for a helpful discussion. This work is supported in part by the European Cooperation in Science and Technology (COST) action MP0905 ``Black Holes in a Violent  Universe" and by the Science and Technology Facilities Council (grant numbers ST/1506029/1 and ST/J000477/1).


\bigskip{}

\end{document}